\documentclass[letterpaper,twocolumn,prl,aps,superscriptaddress,amsmath,amssymb,floatfix]{revtex4-1}
\usepackage{mathptmx}
\usepackage[latin9]{inputenc}
\usepackage{color}
\usepackage{amsmath}
\usepackage{amssymb}
\usepackage{graphicx}
\usepackage{esint}
\usepackage{enumitem}
\usepackage[unicode=true,
 bookmarks=true,bookmarksnumbered=false,bookmarksopen=false,
 breaklinks=false,pdfborder={0 0 1},backref=false,colorlinks=true]
 {hyperref}
\hypersetup{
 linkcolor=magenta,urlcolor=blue,citecolor=blue,pdfstartview={FitH}}

\makeatletter

\pdfpageheight\paperheight
\pdfpagewidth\paperwidth

\usepackage{textcomp}
\usepackage{epstopdf}

\pdfpageheight\paperheight
\pdfpagewidth\paperwidth

\@ifundefined{textcolor}{}{%
 \definecolor{BLACK}{gray}{0}
 \definecolor{WHITE}{gray}{1}
 \definecolor{RED}{rgb}{1,0,0}
 \definecolor{GREEN}{rgb}{0,1,0}
 \definecolor{BLUE}{rgb}{0,0,1}
 \definecolor{CYAN}{cmyk}{1,0,0,0}
 \definecolor{MAGENTA}{cmyk}{0,1,0,0}
 \definecolor{YELLOW}{cmyk}{0,0,1,0}
}

\usepackage{xcolor}
\usepackage{soul}
\definecolor{blue}{rgb}{0,0,1}
\definecolor{red}{rgb}{1,0,0}
\definecolor{green}{rgb}{0,1,0}

\usepackage{soul}

\makeatother

\begin{document}

%\title{Lifting degeneracy of counter-propagating modes in integrated microring resonators without magnetic fields}

\title{Magnetic-free optical mode degeneracy lifting in lithium niobate microring resonators}

\author{Xin-Biao~Xu}
\thanks{These authors contributed equally to this work.}
\affiliation{Laboratory of Quantum Information, University of Science and Technology of China, Hefei 230026, China.}
\affiliation{CAS Center For Excellence in Quantum Information and Quantum Physics, University of Science and Technology of China, Hefei, Anhui 230026, China.}

\author{Zheng-Xu~Zhu}
\thanks{These authors contributed equally to this work.}
\affiliation{Laboratory of Quantum Information, University of Science and
Technology of China, Hefei 230026, China.}
\affiliation{CAS Center For Excellence in Quantum Information and Quantum Physics,
University of Science and Technology of China, Hefei, Anhui 230026,
China.}

\author{Yuan-Hao~Yang}
\affiliation{Laboratory of Quantum Information, University of Science and
Technology of China, Hefei 230026, China.}
\affiliation{CAS Center For Excellence in Quantum Information and Quantum Physics,
University of Science and Technology of China, Hefei, Anhui 230026,
China.}

\author{Jia-Qi~Wang}
\email{wang1221@ustc.edu.cn}
\affiliation{Laboratory of Quantum Information, University of Science and
Technology of China, Hefei 230026, China.}
\affiliation{CAS Center For Excellence in Quantum Information and Quantum Physics,
University of Science and Technology of China, Hefei, Anhui 230026,
China.}

\author{Yu~Zeng}
\affiliation{Laboratory of Quantum Information, University of Science and
Technology of China, Hefei 230026, China.}
\affiliation{CAS Center For Excellence in Quantum Information and Quantum Physics,
University of Science and Technology of China, Hefei, Anhui 230026,
China.}

\author{Jia-Hua~Zou}
\affiliation{Laboratory of Quantum Information, University of Science and
Technology of China, Hefei 230026, China.}
\affiliation{CAS Center For Excellence in Quantum Information and Quantum Physics,
University of Science and Technology of China, Hefei, Anhui 230026,
China.}

\author{Juanjuan~Lu}
\affiliation{School of Information Science and Technology, ShanghaiTech University, Shanghai 201210, China.}

\author{Yan-Lei~Zhang}
\affiliation{Laboratory of Quantum Information, University of Science and
Technology of China, Hefei 230026, China.}
\affiliation{CAS Center For Excellence in Quantum Information and Quantum Physics,
University of Science and Technology of China, Hefei, Anhui 230026,
China.}

\author{Weiting~Wang}
\affiliation{Center for Quantum Information, Institute for Interdisciplinary Information Sciences, Tsinghua University, Beijing 100084, China.}

\author{Guang-Can~Guo}
\affiliation{Laboratory of Quantum Information, University of Science and
Technology of China, Hefei 230026, China.}
\affiliation{CAS Center For Excellence in Quantum Information and Quantum Physics,
University of Science and Technology of China, Hefei, Anhui 230026,
China.}
\affiliation{Hefei National Laboratory, Hefei 230088, China.}

\author{Luyan~Sun}
\email{luyansun@tsinghua.edu.cn}
\affiliation{Center for Quantum Information, Institute for Interdisciplinary Information Sciences, Tsinghua University, Beijing 100084, China.}
\affiliation{Hefei National Laboratory, Hefei 230088, China.}

\author{Chang-Ling~Zou}
\email{clzou321@ustc.edu.cn}
\affiliation{Laboratory of Quantum Information, University of Science and
Technology of China, Hefei 230026, China.}
\affiliation{CAS Center For Excellence in Quantum Information and Quantum Physics,
University of Science and Technology of China, Hefei, Anhui 230026,
China.}
\affiliation{Hefei National Laboratory, Hefei 230088, China.}

%\date{\today}

\begin{abstract}
Breaking time-reversal symmetry in integrated photonics without magnetic fields remains a fundamental challenge. We demonstrate phonon-induced non-reciprocity through direct lifting of forward-backward mode degeneracy in microring resonators. Coherent acousto-optic coupling generates differential AC Stark shifts between counter-propagating fundamental optical modes, eliminating the need for intermodal conversion or complex photonic structures. Simple microwave excitation of integrated piezoelectric transducers provides dynamic control of non-reciprocal response, with experimentally demonstrated mode splitting exceeding twice the optical linewidth. The linear relationship between the splitting and acoustic power enables real-time reconfigurability across a wide range of optical wavelengths. This mechanism requires only simple microring resonators and fundamental optical modes, transforming non-reciprocity from a specialized technique requiring careful modal engineering to a universal, electrically-controlled functionality. Our approach establishes a new paradigm for magnetic-free optical isolation and dynamic topological photonics.
\end{abstract}

\maketitle

\noindent \textit{Introduction.-} Light-matter interactions fundamentally govern electromagnetic energy flow in optical systems, with reciprocity, the symmetry under time reversal, emerging from microscopic reversibility and energy conservation~\cite{Potton2004,Caloz2018,Jalas2013}. Breaking this symmetry to achieve non-reciprocal propagation has profound implications, from protecting laser sources to enabling topological photonic states~\cite{Wang2025,Lu2014,Hafezi2011,Wang2009}. However, creating the necessary asymmetric coupling while preserving optical coherence remains a significant challenge~\cite{Hu2021}. While magneto-optic effects provide strong, broadband non-reciprocity through Faraday rotation, their reliance on materials incompatible with silicon photonics and the need for strong, static magnetic fields limit their dynamic reconfigurability and integration~\cite{Bi2011,Li2024,Shoji2014,Pintus2025}. Magnetic-free alternatives include optomechanical approaches using radiation pressure~\cite{Shen2016,Ruesink2016,Miri2017,Fang2017,Hafezi2012}, nonlinear optical frequency conversion~\cite{Shi2015,Xia2018,Tripathi2024,White2023}, spatio-temporal modulation~\cite{Yu2009a,Lira2012,Estep2014,Sounas2017,Tian2021} and atomic vapor systems~\cite{Zhang2018,Yang2019,Lin2019,Zhang2025,Lu2021,Zhangyl2025}, but each faces critical limitations: optomechanical methods suffer narrow bandwidth, nonlinear frequency conversion requires precise phase matching that restricts operational bandwidth and necessitates complex multi-mode engineering with high optical pump powers, spatio-temporal modulation requires complex multi-stage modulation elements and can only produce a limited and discrete modulation wave vector while atomic systems need specialized cells and temperature control. Consequently, the inherent complexity of managing multiple optical modes, stringent phase-matching conditions, and fabrication tolerances have prevented widespread adoption of these magnetic-free approaches despite their complementary metal-oxide-semiconductor (CMOS) compatibility.

Recently, acousto-optic interactions have emerged as a promising platform for achieving optical nonreciprocity through carefully engineering the coupling between traveling phonons and photons~\cite{Kim2015,Dong2015,Kittlaus2018a,Liu2019,Chen2023,Sohn2021,Kittlaus2021,Sohn2019,Zhou2024}. These schemes exploit the large momentum carried by GHz-frequency acoustic waves to bridge the wavevector mismatch between distinct optical modes, enabling efficient unidirectional conversion between fundamental and higher-order optical modes~\cite{Kittlaus2017,Sohn2018}. Several implementations have successfully demonstrated near-unity conversion efficiency with excellent isolation ratios~\cite{Zhang2024}. However, these intermodal approaches inherit fundamental limitations that restrict their practical deployment. As mode demultiplexers or optical filters are always required to block the converted higher-order modes to improve isolation contrast,  adding complexity and insertion loss~\cite{Sohn2018,Kittlaus2021,Zhou2024}. More critically, phase matching is a fundamental physical requirement. The stringent phase-matching requirement between acoustic and optical modes constrains device operation to specific wavelengths determined by the waveguide geometry, preventing broadband operation or post-fabrication tuning. The need to precisely engineer the dispersion of multiple optical modes while maintaining acoustic coupling further complicates device design and limits scalability.

In this Letter, we demonstrate a fundamentally different approach to acousto-optic non-reciprocity that operates entirely within the fundamental optical mode family. Our mechanism exploits phonon-mediated coherent coupling between clockwise (CW) and counter-clockwise (CCW) propagating modes in microring resonators, where the traveling acoustic wave induces off-resonant coupling that generates propagation direction-dependent AC Stark shifts~\cite{Delone1999}. This off-resonant coupling creates the effect of an effective magnetic field and generates an antisymmetric effect in reversed propagation directions, experiencing differential energy shifts that directly lift their degeneracy. This acoustically induced degeneracy lifting is phenomenologically analogous to the Sagnac effect observed in physically rotating microcavities~\cite{Sarma2015,Ge2015,Maayani2018,Jiao2020}, yet it is realized on a fully stationary integrated platform with electrical reconfigurability. Unlike the inter-modal conversion scheme in Ref.~\cite{Sohn2018}, our approach breaks time-reversal symmetry without mode conversion or auxiliary optical modes. It requires no extra optical degree-of-freedom (neither frequency, polarization, nor spatial domain), providing an elegant and simple approach for realizing non-reciprocity with only fundamental modes and standard suspension-free ring resonators and eliminating the constraints of modal engineering for phase matching. Furthermore, our suspension-free platform offers superior mechanical and thermal robustness, while confined phononic waveguide mode can enhance the acousto-optic interactions. Our work introduces a new paradigm for integrated magnet-free non-reciprocity, offering a scalable and robust pathway toward CMOS compatible isolators and circulators for advanced photonic circuits.

%By operating in the off-resonant regime, the interaction preserves the spatial profile of the optical modes while imparting direction-dependent phase shifts, enabling true non-reciprocal propagation within a single mode family. This mechanism represents a conceptual shift from avoiding backward coupling to embracing symmetric participation of both directions, with the phonon field serving as the symmetry-breaking element that distinguishes forward from backward propagation.

\begin{figure}[t]
\includegraphics[width=1\columnwidth]{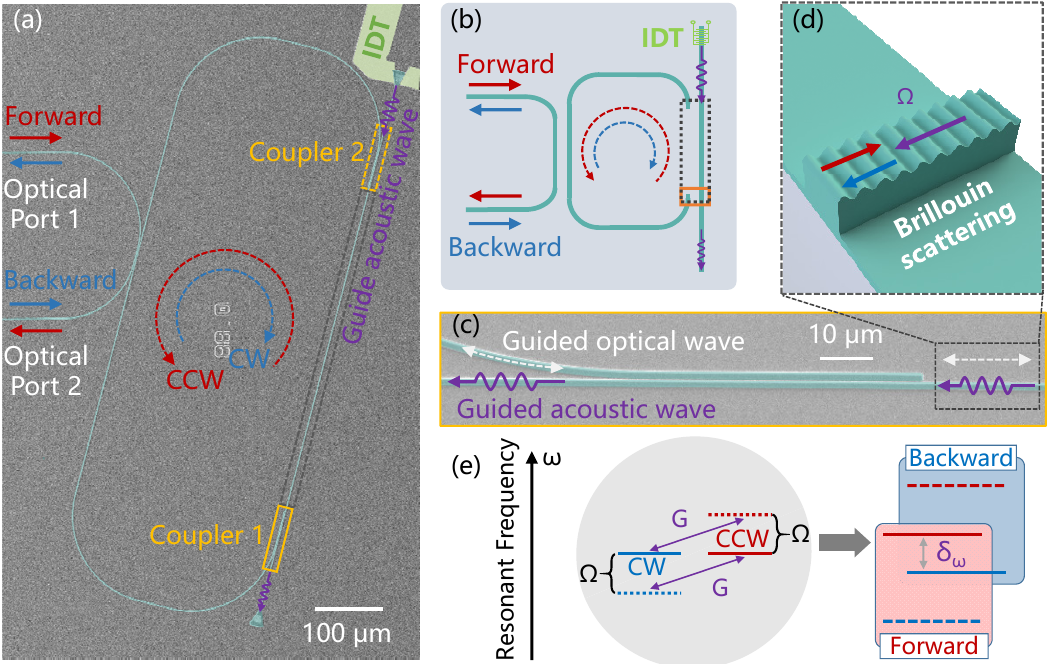}
\caption{\textbf{Phonon-induced lifting of mode degeneracy in microring resonators.} (a) The scanning electron microscope (SEM) image of the device, which includes a splitting microring resonator for the acousto-optic interaction. In the device, a traveling acoustic wave (frequency $\Omega$) couples clockwise (CW) and counter-clockwise (CCW) propagating optical modes through off-resonant interaction, inducing differential AC Stark shifts of the optical modes. IDT: Interdigital Transducer (b) Schematic of the device. (c) The photon-phonon demultiplexing device. The photon modes can couple completely between the waveguides, but waveguide coupling remains negligible for guided phononic modes. (d) The Brillouin interaction in the straight waveguide. (e) Energy level diagram illustrating the phonon-mediated coupling mechanism. The acoustic field creates asymmetric coupling between CW and CCW modes, lifting their degeneracy by $\delta_{\omega}$.
}
\label{Fig1}
\end{figure}

%Our device is a microring resonator composed by separated waveguides.

\noindent \textit{Device and principle.-} Figure~\ref{Fig1}(a) shows the optical microring device we designed for realizing degeneracy lifting, which is made by lithium niobate (LN) ridge waveguide on a sapphire substrate (see Ref.~\cite{sm} for detailed fabrication process). Instead of a single closed-loop waveguide, the optical microring is constructed by a splitting ring structure by coupling a ``C"-shaped bended waveguide and a straight waveguide as depicted in Fig.~\ref{Fig1}(b). By carefully designing the parameters of directional couplers, an effective close-loop for fundamental optical modes can be realized when the power coupling ratio approaches $100\%$. As detailed in Fig.~\ref{Fig1}(b) and (c), the optical coupler structure serves as a photon-phonon demultiplexing device: both phononic and photonic modes can propagate in the straight waveguide~\cite{Yang2023} and be separated completely after the optical coupler as the waveguide tunneling coupling for phononic mode is negligible.
Therefore, the acousto-optics interactions can only occur in the straight waveguide region (the black dashed rectangle), i.e., the straight waveguide segment of the split microring for coherent backward Brillouin interaction between the guided phonon and photon modes~\cite{sm}, as illustrated in Fig.~\ref{Fig1}(d). The principle and demonstration of Brillouin interaction between guided phonons (around 9\,GHz) and photons (around telecom wavelength band) in a single LN waveguide can be found in Ref.~\cite{Yang2023,Rodrigues2025,Ye2025,Yu2025,Nie2025}. Therefore, the microring is coupled through two optical ports and two acoustic ports [Fig.~\ref{Fig1}(a)]. The optical signal can probe either forward (red arrow) or backward (blue arrow) direction, and the acoustic wave (purple arrow) is excited by an interdigital transducer (IDT)~\cite{sm}.

Under the acoustic pumping, the key physics of acoustically-induced lifting of mode degeneracy in our device can be described by the Hamiltonian~\cite{sm}
\begin{align}
H/\hbar  = & \omega_{0}(a_{\mathrm{cw}}^{\dagger}a_{\mathrm{cw}}+a_{\mathrm{ccw}}^{\dagger}a_{\mathrm{ccw}})\label{eq:1}\\
\nonumber
&+(G\hat{a}_{\mathrm{ccw}}^{\dagger}\hat{a}_{\mathrm{cw}}e^{i\Omega t}+G^{\ast}\hat{a}_{\mathrm{cw}}^{\dagger}\hat{a}_{\mathrm{ccw}}e^{-i\Omega t}),
\end{align}
where $\omega_{0}$ is the optical mode frequency, $a$ and $a^{\dagger}$ are bosonic operators with subscripts denoting CW and CCW modes, $G$ is the acoustic-stimulated backward Brillouin coupling strength between CW and CCW modes, with input acoustic wave frequency of $\Omega$. The corresponding energy diagram is illustrated in Fig.~\ref{Fig1}(e), where the resonant frequency of initial CW and CCW modes are degenerate as the red and blue solid lines shown in the gray shadow region.

When the acoustic pumping is on, as indicated by the purple arrows, the CW and CCW modes are coupled to each other with a detuning frequency $\Omega$. For the forward optical signal at frequency $\omega$ probing the microring, it simultaneously excites the CCW mode at frequency $\omega$ and the CW mode at frequency $\omega+\Omega$. The CW photons have higher energy as they are generated by the CCW photons that absorb the phonons in the backward Brillouin interaction region [Fig.~\ref{Fig1}(c)]. Consequently, the CCW mode is off-resonantly coupled with the CW mode, and gives rise to mode hybridization, which modifies the eigen-frequency of the CCW mode from $\omega_0$ to $\omega_0+(\sqrt{\left|G\right|^2+\Omega^2/4}-\Omega/2)$ as the energy levels of group ``Forward" shown in Fig.~\ref{Fig1}(e).

At large detuning limit $\Omega\gg G$, the situation is equivalent to the red-detuned coupling to another energy level in atomic physics, leading to an AC Stark shift on the CCW mode as
\begin{equation}
    \Delta_{\mathrm{ccw}}^{\mathrm{ac}}=\sqrt{\left|G\right|^2+\Omega^2/4}-\Omega/2\approx\frac{\left|G\right|^2}{\Omega}.
    \label{eq:3}
\end{equation}

Similarly, the probe from the backward direction excites the CW mode at frequency $\omega$, while converting CCW photon at frequency $\omega-\Omega$. Therefore, an equivalent blue-detuned coupling between the CW mode and the CCW mode is simulated by the acoustic wave, accompanied by an AC Stark shift on the CW mode $\Delta_{\mathrm{cw}}^{\mathrm{ac}}=-\Delta_{\mathrm{ccw}}^{\mathrm{ac}}$. Then, the resulting normal modes have the energy levels of group ``Backward", as shown in Fig.~\ref{Fig1}(e). Finally, the mode splitting between the CW and CCW modes is $\delta_{\omega}=2\Delta_{\mathrm{cw}}^{\mathrm{ac}}$. It is interesting to note that two additional dashed energy levels can also be probed from both forward and backward directions.

The two optical modes with opposite orbit angular momenta exhibit opposite frequency shifts under the same external acoustic pump field, and thus the degeneracy of the CW and CCW modes is lifted. This effect resembles the Zeeman effect for electron spins in an external magnetic field, where opposite spin angular momenta possess opposite energy shifts. The external acoustic field breaks the time-reversal symmetry, and the probes from forward and backward exhibit different resonant frequencies, leading to optical non-reciprocity. It is worth noting that our system requires no extra optical mode families, and the acoustic interaction shows exact antisymmetry between the CW and CCW modes. This mechanism is significantly different from all previous approaches to realize non-magnetic non-reciprocity.

\begin{figure}[t]
\includegraphics[width=1\columnwidth]{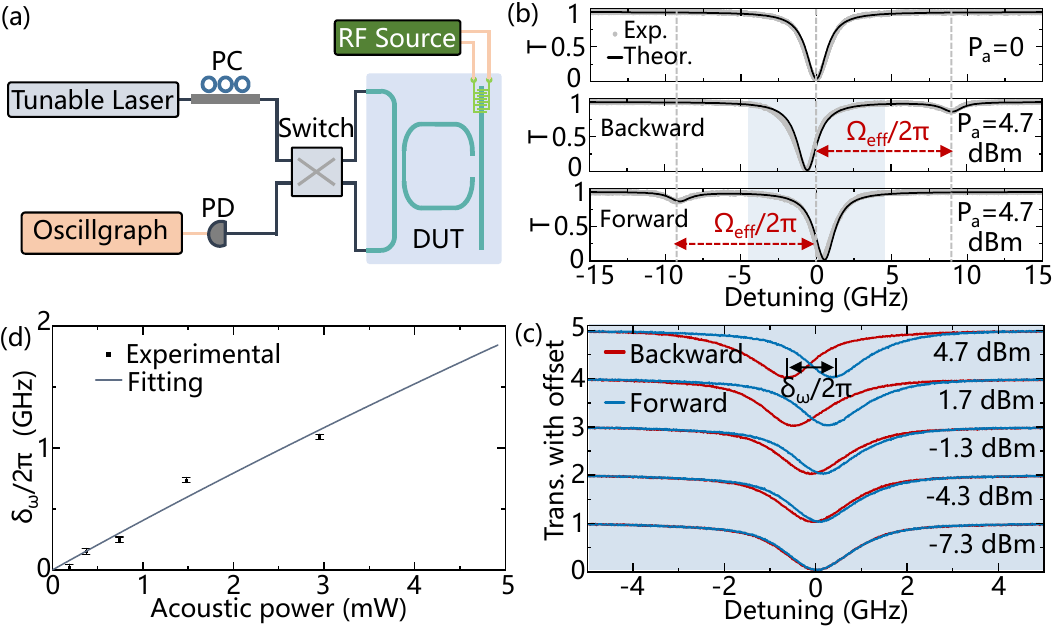}
\caption{\textbf{Experimental characterization of the microcavity transmission degeneration.} (a) Experimental setup for measuring transmission spectrum degeneration. PC: fiber polarization controller; Switch: optical path switch; PD: optical detector; RF source: radio frequency source, DUT: device under test. (b) The experimentally measured microcavity transmission spectrum degeneration. The gray dots are the experimental results and the black lines are the theoretical fit. The upper panel shows the microcavity transmission spectrum when the RF driver is off. The middle and bottom panels show the transmission spectrum with backward and forward optical inputs, respectively, with an acoustic drive power of 4.7\,dBm. (c) Enlarged plots of the biased transmission spectra of the blue region in (b) for different RF drive powers. The red and blue line are the backward and forward transmission spectrum, respectively. $\delta_{\omega}/2\pi$ indicates the resonance frequency splitting of the transmission spectrum. (d) $\delta_{\omega}$ versus RF drive power, where the black data points are experimental data with error bars indicating 10 times the standard deviation. The solid line is the fit to the experimental data.}
\label{Fig2}
\end{figure}

\begin{figure}[t]
\includegraphics[width=1\columnwidth]{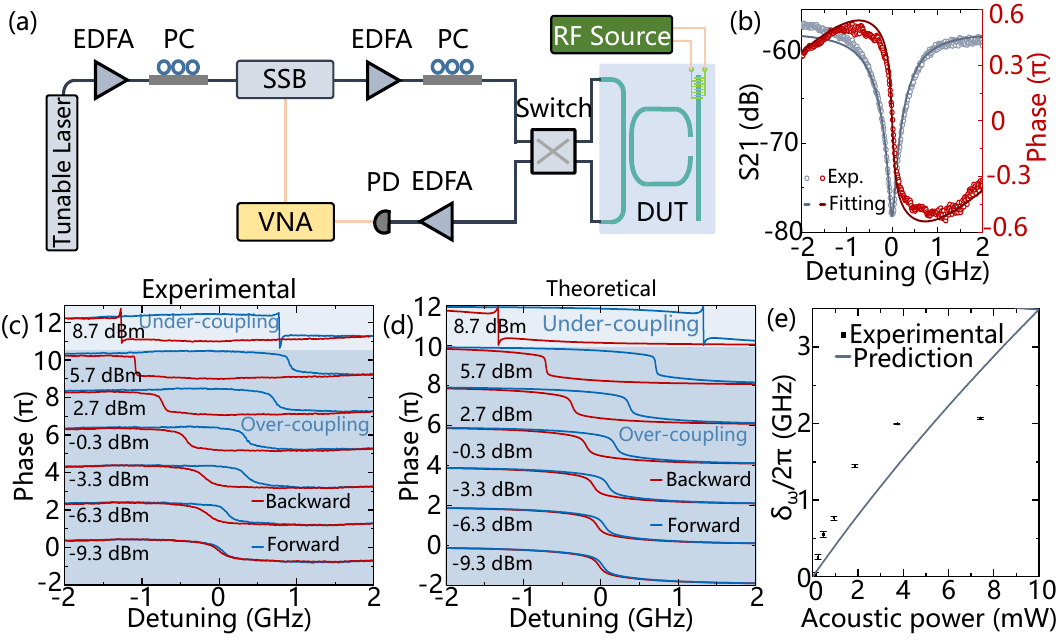}
\caption{\textbf{Experimental characterization of the microcavity phase spectrum degeneration.} (a) Experimental setup for measuring phase spectrum degeneration. EDFA: erbium-doped fiber amplifier; VNA: Vector Network Analyzer; SSB: single-sideband modulator; DUT: device under test. (b) The transmission and phase spectrum of the microring with the acoustic power of -9.3\,dBm, respectively. The doted curves are the experimental data and the solid lines are the fit. (c) The experimental phase spectrum of the microring at different acoustic drive power. (d) The theoretical phase
spectrum calculated with the fitted $\alpha$ value from Fig.~\ref{Fig2}(d). (e) $\delta_{\omega}$ versus acoustic drive power. The doted data are the experimental results obtained according to the phase spectrum with error bars indicating 10 times the standard deviation. The solid curve is the theoretical prediction with the $\ensuremath{\alpha}$ value obtained from Fig.~\ref{Fig2}(d).
}
\label{Fig3}
\end{figure}

\begin{figure*}[t]
\includegraphics[width=0.9\textwidth]{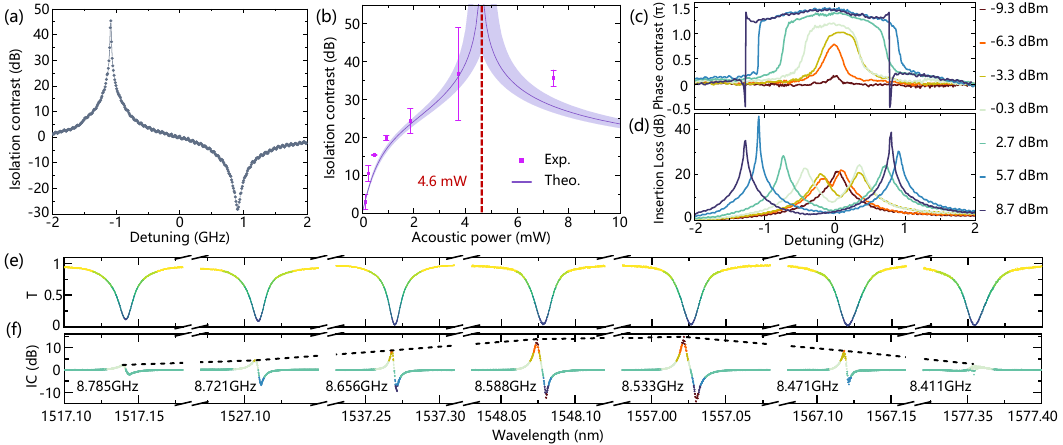}
\caption{\textbf{Experimental characterization of non-reciprocal transmission.} (a) The isolation contrast (IC) of the microring when acoustic drive power is $5.7\,\mathrm{dBm}$. The maximum obtained isolation contrast is about 45 dB. (b) The maximum isolation contrast measured at different acoustic drive powers. The dotted data are the experimental results. The purple solid line is the theoretical result calculated according to the average value of the fitted $\alpha$, and the light purple area is the interval of the theoretical results calculated according to the error bounds of $\alpha$. (c) and (d) The experimental phase contrast between forward and backward transmissions and the corresponding insertion loss of the device, respectively. (e) and (f) Typical cavity resonant modes with the corresponding isolation contrast we measured. The frequencies shown in (f) are the corresponding RF drive frequencies of each optical mode. The black dash envelope in (f) shows the limited bandwidth of the isolation contrast attributed to the bandwidth of IDT.
}
\label{Fig4}
\end{figure*}

\noindent \textit{The degeneracy lifting in transmission spectra.-} The degeneracy lifting effect is experimentally validated by probing the transmission spectra from opposite directions. The experimental setup is shown in Fig.~\ref{Fig2} (a), where a telecom tunable laser is used as optical signal, and the output light is collected and detected by a photo-detector. An optical switch is employed to alternate the forward and backward optical probes of the microring resonator, with the polarization of the optical input optimized by a polarization controller for exciting the transverse electric (TE) modes in the waveguide. The on-chip acoustic pumping power ($P_{\mathrm{a}}$) is controlled by the power of the RF source, with an RF-acoustic wave conversion efficiency of about $10\%$~\cite{sm}.

Figure~\ref{Fig2} (b) shows the typical transmission spectra with an optical wavelength at around $1561\,\mathrm{nm}$ and the corresponding fitting results. When acoustic pumping is off ($P_{\mathrm{a}}=0$), as a result of optical reciprocity, the CW and CCW modes show identical spectra with a single Lorentzian resonance dip, i.e., the two modes have the same resonant frequency ($\omega_0$), intrinsic loss, and the external coupling rate to the bus waveguide. When acoustic pump is applied ($P_{\mathrm{a}}=4.7\,\mathrm{dBm}$ and $\Omega/2\pi=8.533\,\mathrm{GHz}$), the forward and backward spectra become different, but the two spectra show mirror symmetry with respect to $\omega_0$. Furthermore, there is an additional dip with a efficitivy detuning $\Omega_\mathrm{eff}=\Omega/2+\sqrt{\Omega^2/4+\left|G\right|^2})$ in the backward spectrum, which agrees with our theoretical prediction [Fig.~\ref{Fig1}(d)]. The lifted degeneracy is systematically investigated under varying $P_{\mathrm{a}}$, as summarized in Fig.~\ref{Fig2}(c). The resulting frequency splitting $\delta_{\omega}$ at different $P_{\mathrm{a}}$ is extracted and plotted in Fig.~\ref{Fig2}(d). According to our theory, the spitting can be solved through $\delta_{\omega}=\sqrt{4\left|G\right|^2+\Omega^2}-\Omega$, where the interaction strength is determined by the acoustic pump power as $G^2=\alpha P_{\mathrm{a}}$. The scaling factor $\alpha=1774\pm171\,\mathrm{GHz^{2}/W}$ is obtained by fitting the experimental results.

The lifted degeneracy is further verified by characterizing the transmittance phase. In contrast to other non-reciprocal mechanisms that employ direction-dependent conversion between optical modes, our approach induces the antisymmetric frequency shifts to both CW and CCW modes, thus enabling antisymmetric phase responses when probed from opposite directions. Using the setup in Fig.~\ref{Fig3}(a),  the amplitude and phase responses of the optical signal are characterized by a vector network analyzer (VNA), with the optical signal generated from a local oscillator (LO) laser passing through a single-sideband modulator. The resulting beat signal between the single-sideband and the LO is sent to the VNA for demodulation. For instance, Fig.~\ref{Fig3}(b) presents the intensity and phase spectra of the CW optical mode under an acoustic pumping power of $-9.3\,\mathrm{dBm}$ with the fitted cavity amplitude dissipation rates being $\kappa_{0}/2\pi\approx0.32\,\mathrm{GHz}$ and $\kappa_{1}/2\pi\approx0.38\,\mathrm{GHz}$. Figure~\ref{Fig3}(c) shows the measured forward and backward phase spectra at various $P_{\mathrm{a}}$, while Fig.~\ref{Fig3}(d) provides the theoretical prediction with the extracted $\alpha$ in Fig.~\ref{Fig2}(d). We can see that when acoustic drive power is 5.7~dBm, the mode splitting is about $2~\mathrm{GHz}$, exceeding twice of the
optical linewidth. Although experimental imperfections, such as the frequency instability of the laser and environmental thermal fluctuations induced resonance frequencies drift, both experimental and theoretical results confirm the trends of increasing opposite frequency shift when increasing $P_\mathrm{a}$. Particularly, as the effective intracavity dissipation rate $\kappa_{0}$ are positively related to the coupling strength $G$~\cite{sm}, the phase spectra reveal the transition from the over-coupling to under-coupling when $P_{\mathrm{a}}=5.7\,\mathrm{dBm}$. In Fig.~\ref{Fig3}(e), the frequency splitting extracted from the experimental phase spectra shows a saturation-like effect, this is actually caused by environmental thermal fluctuations. Overall, the splitting remains consistent with the theoretical prediction with the scaling factor $\alpha$ obtained in Fig.~\ref{Fig2}(d). Based on the $\delta_{\omega}$ function, the slope of the splitting change is determined by $d\delta_\omega / dP_a = 2\alpha / \sqrt{\Omega^2 + 4\alpha P_a}$. In the weak coupling regime ($\Omega^2 \gg 4\alpha P_a$), the slope can be approximated as $2\alpha/\Omega$. It clarifies that a larger $\alpha$ or a smaller coupling detuning $\Omega$ leads to a higher tuning efficiency~\cite{sm}.

\noindent \textit{Broadband Nonreciprocity.-} The optical non-reciprocity due to the lifted degeneracy of the optical resonant modes is characterized in Fig.~\ref{Fig4}. The optical isolation ratio spectra of the system, i.e., the transmittance ratio between the forward and backward probe directions, are measured and shown in Fig.~\ref{Fig4}(a). A maximum isolation of 45\,dB under an acoustic drive power of $P_{\mathrm{a}}=5.7\,\mathrm{dBm}$ is achieved. Note that this power is critical and the optimal isolation contrast is achieved when the critical coupling condition is achieved. Figure~\ref{Fig4}(b) shows the optical isolation contrast of the system under different $P_{\mathrm{a}}$. The dots are obtained from experimental results, while the black solid line is the theoretical prediction with the given $\alpha$ in Fig.~\ref{Fig2}(d) and cavity dissipation rate
 obtained in Fig.~\ref{Fig3}(b). The experimental and theoretical results are in good agreement. Theory predicts that maximum isolation can be obtained with $P_{\mathrm{a}}$ of only around $4.6\,\mathrm{mW}$, showing a low power consumption that has great potential in practical applications.

Our device functions not only as an isolator but also as an optical gyrator~\cite{Orsel2025}, which imparts different phases to forward and backward propagating light. Figure~\ref{Fig4}(c) plots the phase contrast of the opposit transmission spectra, while the corresponding insertion losses are shown in Fig.~\ref{Fig4}(d). It is found that the phase contrast is negligible (no gyration) when the modes are degenerate with $P_{\mathrm{a}}\approx0$, and its value is largest at resonant frequency point $\omega_0$ and increases to $2\pi$ with increasing $P_{\mathrm{a}}$. Remarkably, by applying just $P_{\mathrm{a}}=-0.3\,\mathrm{dBm}$ acoustic power, we achieve a $\pi$-phase gyration with an insertion loss of about $8.2\,\mathrm{dB}$, which is particularly valuable for constructing other functional non-recprical optical devices through interference, such as multi-port circulations.

The Brillouin interaction in straight waveguides exhibits naturally a large bandwidth~\cite{Yang2023,Shin2013,Eggleton2019}. In our device it enables the optical bandwidth around 1\,nm for a given acoustic pump frequency $\Omega$. Additionally, we can always shift this window of the optical signal by changing $\Omega$, and thus realize degeneracy lifting and non-reciprocal transmission for a given optical mode over a large wavelength range. As shown in Figs.~\ref{Fig4}(e) and \ref{Fig4}(f), we demonstrate the non-reciprocal response across seven different resonances spanning 60\,nm (1517~nm-1577~nm), by adjusting the RF drive frequency at a fixed power of $15\,\mathrm{dBm}$. Since the power of acoustic pumping is limited by the bandwidth of IDT, the induced frequency splitting and isolation contrast decreases when $\Omega$ deviates from IDT's central frequency,as shown by the black dashed line in Fig.~\ref{Fig4}(f). It indicates that an even larger wavelength bandwidth can be achieved with optimized IDTs designs. Furthermore, combining our acoustic control with established optical resonant frequency tuning methods (thermo-optic or electro-optic) would enable non-reciprocal operation at arbitrary wavelengths in the 60\,nm window, making this a truly universal approach for on-chip optical devices.

\noindent \textit{Conclusion.-} Under an external acoustic bias, we have demonstrated a mechanism for lifting the forward and backward mode degeneracy in an optical microring resonator through a manner similar to the Zeeman effect acting on spin states. The time-reversal symmetry is broken through the electric drive-stimulated backward Brillouin scattering, without requiring optical pumping, extra optical modes, or complex engineering for phase matching. This effect provides a novel approach to realizing optical non-reciprocity by allowing both counter-propagating modes to participate in the acousto-optic interaction while experiencing opposite AC Stark shifts. Looking forward, our results establish acoustically-induced degeneracy lifting as a versatile platform for programmable non-reciprocal photonic devices, including high performance isolators and gyrators, enabling reconfigurable topological photonic circuits.

\smallskip{}
\noindent\textit{Acknowledgments---}
This work was funded by the Quantum Science and Technology-National Science and Technology Major Project (Grant Nos.~2024ZD0301500 and 2021ZD0300200) and the National Natural Science Foundation of China (Grant Nos.~92265210, 92265108, 123B2068, 12550006, 92365301, 92565301, 12474498, 12374361, 12293053, 12404567, and 92165209). We also acknowledge the support from the Fundamental Research Funds for the Central Universities and USTC Research Funds of the Double First-Class Initiative. CLZ was also supported by Beijing National Laboratory for Condensed Matter Physics (2024BNLCMPKF007). The numerical calculations in this paper were performed on the supercomputing system in the Supercomputing Center of University of Science and Technology of China. This work was partially carried out at the USTC Center for Micro and Nanoscale Research and Fabrication.

\smallskip
\noindent\textit{Data availability---}
The data that support the findings of this study are openly available \cite{data_figshare}, embargo periods may apply.

\end{document}